\documentclass[10pt,twocolumn,english,aps,prl,nofootinbib]{revtex4}

\usepackage[utf8]{inputenc}
\usepackage{graphicx}
\usepackage{amsmath}
\usepackage{amssymb}
\usepackage{amsfonts}
\usepackage{lscape}
\usepackage{hyperref}
\usepackage{url}
\usepackage{epsfig}
\usepackage{color}
\usepackage{bm}
\usepackage{tabularx}
\usepackage{braket}
\newcommand{\be}{\begin{equation}}
\newcommand{\ee}{\end{equation}}
\newcommand{\ba}{\begin{eqnarray}}
\newcommand{\ea}{\end{eqnarray}}

\renewcommand{\[}{\begin{equation}}
\renewcommand{\]}{\end{equation}}

\begin{document}

\thispagestyle{empty}

\title{Long-range enhanced mutual information from inflation}

\author{Lloren\c{c} Espinosa-Portal\'es}\email[]{llorenc.espinosa@uam.es}
\author{Juan Garc\'ia-Bellido}\email[]{juan.garciabellido@uam.es}

\affiliation{Instituto de F\'isica Te\'orica UAM-CSIC, Universidad Auton\'oma de Madrid,
Cantoblanco, 28049 Madrid, Spain}

\date{\today}

\begin{abstract}
The quantum origin of cosmological primordial perturbations is a cornerstone framework in the interplay between gravity and quantum physics. In this paper we study the mutual information between two spatial regions in a radiation-dominated universe filled by a curvature perturbation field in a squeezed state. We find an enhancement with respect to the usual mutual information of the Minkowski vacuum due to momentum modes affected by particle production during inflation. This result supports our previous claim of the existence of long-range correlations between Primordial Black Holes (PBH) at formation during the radiation era.
\end{abstract}
\maketitle

\section{I. Introduction}

Entropy and information play a key role in our understanding of physics. They are important properties of quantum states and are useful in describing correlations. They are thought to be a bridge between classical gravity and an underlying quantum theory of gravity.

The study of entropy and information applied to Black Hole physics is a fruitful field of research. The introduction of Bekenstein entropy~\cite{Bekenstein:1973ur} was followed by the discovery of the area law of entanglement entropy~\cite{Bombelli:1986rw, Srednicki:1993im}. The link between these two concepts added quantum information to the already successful crossover between gravity and quantum field theory.

Cosmology also profits from this interplay between gravity and quantum physics. The idea of inflation introduced a quantum origin of primordial perturbations~\cite{Mukhanov:1990me}. This was needed in order to explain the power spectra of the Cosmic Microwave Background (CMB) and some features of the Large-Scale Structure (LSS) of the universe. Less known alternatives to inflation also explain power spectra by means of quantum fluctuations~\cite{Brandenberger:2018wbg}. Even though quantum fluctuations classicalize in the sense that their observable features appear classical~\cite{Albrecht:1992kf,Prokopec:1992ia, Polarski:1995jg}, their quantum origin is still relevant. For instance, the study of the entropy of cosmological perturbations in momentum space has long been considered~\cite{Brandenberger:1992jh} and has recently been extended to include non-linear interactions~\cite{Brahma:2020zpk}. In a more general sense, it has also been a matter of recent work the study of the universe as a storage of quantum information in gravitational d.o.f., which could in turn leave an imprint on primordial perturbations~\cite{Dvali:2018ytn}.

In a previous work we investigated the entanglement entropy of the primordial curvature perturbation field in a radiation-dominated universe~\cite{Espinosa-Portales:2019peb}. We found that there are UV-finite contributions to the entanglement entropy. These signal contributions from long-range correlations. They are a natural consequence of the stretching of quantum fluctuations during inflation. We argued that entangled Primordial Black Holes (PBH) could be formed by gravitational collapse of entangled perturbations during the radiation era.

Here we take a step further by computing the mutual information between disjoint regions in a radiation-dominated universe filled with a curvature perturbation field. The mutual information quantifies both classical and quantum correlations between these regions. We find that it is linked to the primordial power spectrum and thus enhanced by inflation due to the stretching of quantum fluctuations. 

This paper is organized as follows. In section II we review the concepts of entropy and information associated to quantum states. In section III we describe the squeezed state of the scalar perturbation field in terms of 2-point correlation functions, whose behavior we analyze. In section IV we review the well-known formalism that connects correlation functions and entanglement of local d.o.f. In section V we adapt an existing perturbative approach for the computation of the mutual information to the particular squeezed state of the radiation field and obtain a closed-form expression for it. In section VI we discuss some implications for the cosmological evolution and we finish with conclusions in section VII.


\section{II. Entropy and information}

We provide here a brief review of the concept of the entropy associated to a quantum state and its link to information. Given a quantum state described by its density matrix $\rho$ one defines its von Neumann entropy as
\begin{equation}
	S = - \textrm{Tr} \left(\rho \log \rho \right)\,. 
\end{equation}
This satisfies the simple but important property
\begin{equation}
\begin{aligned}
	&S = 0 \quad \textrm{for} \quad \rho \quad \textrm{a pure state}\\
	&S > 0 \quad \textrm{for} \quad \rho \quad \textrm{a mixed state}\,.
\end{aligned}
\end{equation}

If $\rho$ describes the state of a system with several d.o.f., for instance two complementary subsystems $A$ and $B$, we can ask the same questions regarding one of its reduced density matrices $\rho_A = \textrm{Tr}_B \rho$ 
\begin{equation}
	S_A = -\textrm{Tr} \left(\rho_A \log \rho_A\right)\,. 
\end{equation}
This is the von Neumann entropy of the state $\rho_A$. If $\rho$ is a pure state, then the subsystem $A$ is in a mixed state only if its entangled with $B$. Then $S_A = S_B$ and is called the entanglement entropy.

Let us consider now a multipartite system with possibly infinite d.o.f., as it is the case of a quantum field. Then $A$ and $B$ need not be complementary and one can ask what is the entropy of the subsystem $A \cup B$ and it turns out to be given by
\begin{equation}
	S_{A\cup B} = S_A + S_B - I(A,B)\,.
\end{equation}

$I(A,B)$ is defined as the mutual information between $A$ and $B$ and is the key object of study of this paper. In the case of a quantum field consisting of  local continuous d.o.f., subsystems correspond to local d.o.f. restricted to spatial regions. The mutual information is a measure of total (classical and quantum) correlations between disjoint regions $A$ and $B$. It satisfies two important properties:

\begin{itemize}
    \item Non-negativity
    \begin{equation}I(A,B) \ge 0 \,. \end{equation}
    \item Symmetry
    \begin{equation}I(A,B) = I(B,A) \,. \end{equation}
\end{itemize}

The mutual information between two regions for a scalar field in the Minkowski vacuum is a rapidly decaying function of the distance $r$. For instance, for two spheres of radius $R_1$ and $R_2$ and $R_1,R_2 \ll r$ one finds the expression \cite{Shiba:2010dy, Shiba:2012np}
\begin{equation}
	I(A,B) \simeq \frac{1}{4} \frac{R_1^2 R_2^2}{r^4}\,,
\end{equation}
which becomes quickly irrelevant. We will see in the course of this paper how this quantity is enhanced thanks to particle production (or, equivalently, stretching of quantum fluctuations) during inflation. Indeed, this same quantity for a scalar field in the squeezed state resulting from an inflationary period lasting from conformal time $\eta_0$ to $\eta_{\rm end}$ and evaluated at super-horizon scales at conformal time $\eta$ during the radiation era is given by
\begin{equation}
	I(A,B) \simeq \frac{1}{16} \frac{R_1^2 R_2^2}{\eta_{\rm end}^4} \left(\frac{\eta}{\eta_{\rm end}} \right)^4 \left[1-\gamma+\log\left(\frac{-\eta_0}{r}\right)\right]^2\,,
\end{equation}
where $\gamma \simeq 0.577216...$ is the Euler-Mascheroni constant. This much slower decay signals long-range correlations between these disjoint regions and is the main result of our paper. It is also a natural result: due to inflation distant regions were causally connected in the past. Enhanced mutual information is intuitively connected with the main dynamical prediction of inflation: an homogeneous and isotropic universe with a nearly scale-invariant spectrum of curvature perturbations.


\section{III. The quantum state after inflation}

Consider a FLRW universe with linear perturbations of its geometry and matter content. In the longitudinal gauge, its metric takes the following form \cite{Mukhanov:2005sc, GarciaBellido:1999ys}:

\begin{equation}
    ds^2 = a^2(\eta) \left[\left(1+2\Phi\right)d\eta^2 - \left(1-2\Phi\right)d\vec{x}^2 \right]\,.
\end{equation}

Where $a$ is the scale factor, $\eta$ is the conformal time, $\vec{x}$ is the set of comoving spatial coordinates and $\Phi$ is the gauge-invariant gravitational potential. Notice that all coordinates are dimensionless and only the scale factor $a$ keeps track of physical dimensions. During inflation the background evolution is dominated by a scalar field $\varphi(\eta)$, which also has linear gauge-invariant perturbations $\delta\varphi (\eta, \vec{x})$. One can fully characterize primordial scalar perturbations by means of the Mukhanov-Sasaki variable \cite{Mukhanov:2005sc, GarciaBellido:1999ys, Martin:2012pea}:

\begin{equation}
    v(\eta, \vec{x}) = a(\eta) \left( \delta\varphi + \varphi' \frac{\Phi}{\mathcal{H}} \right)\, ,
\end{equation}

where $'$ denotes a derivative with respect to conformal time $\eta$ and $\mathcal{H} = a'/a$. The origin of these perturbations its traced back to quantum fluctuations that are stretched out during inflation, which drives growth (i.e. particle creation) when modes become super-horizon. The Mukhanov-Sasaki variable is directly connected to gauge-invariant metric curvature perturbations via the relation:

\begin{equation}
v(\eta, \vec{x}) = \frac{\zeta(\eta, \vec{x})}{z}\, ,  
\end{equation}

with $z$ being mainly related to the scale factor:

\begin{equation}
    z = a \frac{\varphi'}{\mathcal{H}} \, .
\end{equation}

The dynamics of of the Mukhanov-Sasaki variable is derived from a perturbation of the action:

\begin{equation}\label{eq:action1}
\delta S = \frac{1}{2} \int d^4 x \left( (v')^2 - c_s^2 \delta^{ij}\partial_i v \partial_j v + \frac{z''}{z} v^2 \right) \, ,
\end{equation}

where $c_s$ is the speed of sound, which takes values $c_s = 1$ during inflation and $c_s= 1/\sqrt{3}$ during the radiation era.  The corresponding equation of motion for the Fourier modes $v_k (\eta) = \int d^3x e^{i\vec{k}\cdot \vec{x}} v(\eta,\vec{x})$ is:

\begin{equation}\label{eq:vz}
    v'' + \left(c_s^2 k^2 - \frac{z''}{z} \right) v = 0 \,,
\end{equation}

which is the equation of motion of a harmonic oscillator with time-dependent mass. Thus, whenever $c_s^2 k^2 < z''/z$, particle creation can occur. 

At the beginning of inflation, the perturbation field is assumed to be in the Bunch-Davies vacuum, i.e. mode functions behave as plane waves in the distant past \cite{Mukhanov:2007zz}. Then these modes evolve and are put in a squeezed state after they become super-horizon. For each momentum mode $k$, the state is described by a squeezing parameter $\tau_k$ and angle $\delta_k$. This time-evolution is due to the $z''/z$ term in the e.o.m. (\ref{eq:vz}).

We will refer mostly to curvature perturbations, but our conclusions can be extended to primordial gravitational waves as well, since they have effectively the same dynamics.

The time-evolution of the quantum state for general inflationary models is more practically obtained after performing a canonical transformation of the Hamiltonian obtained from the action (\ref{eq:action1}). This is equivalent to the addition of a total derivative to the action in order to get:
\begin{equation}
    \delta S = \frac{1}{2} \int d^4 x \left( (v')^2 - c_s^2 \delta^{ij}\partial_i  v \partial_j v  -2 \frac{z'}{z}v v' + \left(\frac{z'}{z}\right)^2v^2 \right)
\end{equation}

Of course, from this action one gets the same equation of motion (\ref{eq:vz}). The canonical momentum and the Hamiltonian are given by
\begin{equation}\label{eq:H}
\begin{aligned}
    &\pi = v' - \frac{a'}{a}v\,,\\
    &H = \frac{1}{2} \int d^3 x \left( \pi^2 + c_s^2 \delta^{ij}\partial_i  v \partial_j v + 2 \frac{z'}{z} v \pi \right)\,.
\end{aligned}
\end{equation}

We will use this Hamiltonian for the rest of the paper. For a detailed discussion of the two Hamiltonians that can equivalently describe the time-evolution of primordial perturbations and the canonical transformation that relates them, we refer the reader to Refs.~\cite{Martin:2007bw, Martin:2012pea}.

In our previous work, we described this state with the bracket formalism. Here it will be more useful to consider its 1- and 2-point correlation functions, which determine any Gaussian state. Not only do we know the squeezed state to be a Gaussian state, but it is a general result that Gaussian states remain Gaussian if the Hamiltonian that drives their evolution is bilinear \cite{Schumaker}. This statement is true regardless of whether the Hamiltonian conserves particle number or not.

Inflation is succeeded by the radiation era. Recall that $\eta \in (-\infty,0)$ for eternal inflation or dS and $\eta \in (0,\infty)$ for an eternal radiation era. Instead, we will consider that inflation starts at $\eta_0 < 0$ and finishes at $\eta_{\rm end} < 0 $ and then the radiation era starts at $-\eta_{\rm end}$. The details of the matching between $\eta_{\rm end}$ and $-\eta_{\rm end}$ depend on the reheating scenario, but have little effect on curvature perturbations. Nevertheless, the mode functions of the radiation era depend of course on the boundary conditions imposed at $-\eta_{\rm end}$. First, we will obtain the correlation functions in quasi de-Sitter inflation by obtaining the time evolution of the mode functions and then generalize them by applying known results in the squeezing formalism.



\subsection{Correlation functions in quasi de-Sitter}

During quasi de-Sitter inflation $c_s^2 = 1$, $a = -1/(H\eta)$ and therefore $z''/z = 2/\eta^2$. In this scenario, the mode functions of the Bunch-Davies vacuum have a simple form:
\begin{equation}
	v^i_k = \frac{(k |\eta|+i)}{ \sqrt{2} \,k^{3/2} |\eta|} e^{i k |\eta|}\, .
\end{equation}

From them one computes the mode functions for the canonical momentum
\begin{equation}
	\pi^i_k(\eta) = v^{i'}_k(\eta)-\frac{a'}{a} v^i_k(\eta) = \frac{i\sqrt{k}}{\sqrt{2}} e^{ik|\eta|}\,.
\end{equation}

Mode functions allow us to build the mode expansions of the quantum field and its canonical momentum
\begin{equation}
\begin{aligned}
	&v^i(\eta,\vec{x}) = \int \frac{d^3 k}{(2\pi)^{3/2}} \left(e^{i\vec{k}\vec{x}} v_k^{i*}(\eta) \hat{a}_k + e^{-i\vec{k}\vec{x}} v^i_k(\eta) \hat{a}_k^{\dagger} \right) \,,\\
	&\pi^i(\eta,\vec{x}) = \int \frac{d^3 k}{(2\pi)^{3/2}} \left(e^{i\vec{k}\vec{x}} \pi_k^{i*}(\eta) \hat{a}_k + e^{-i\vec{k}\vec{x}} \pi^i_k(\eta) \hat{a}_k^{\dagger} \right)\,.
\end{aligned}
\end{equation}

Now we can compute the correlation functions that characterize the Bunch-Davies vacuum during quasi de-Sitter inflation:

\begin{itemize}
	\item 1-point correlation functions
	\begin{equation}
		\braket{v^i(x,\eta)} = \braket{\pi^i(x,\eta)} = 0\,.
	\end{equation}

\item 2-point correlation functions
	\begin{equation}
	\begin{aligned}
		&\braket{v^i(\eta, \vec{x})v^i(\eta, \vec{y})} = \int \frac{d^3k}{(2\pi)^3} \frac{1}{2k} \left(1 + \frac{1}{k^2 \eta^2} \right) e^{-i \vec{k} (\vec{x}-\vec{y})} \,, \\
		&\braket{\pi^i(\eta, \vec{x}) \pi^i(\eta, \vec{y})}  = \int \frac{d^3k}{(2\pi)^3} \frac{k}{2} e^{-i \vec{k} (\vec{x}-\vec{y})}\,, \\
		&\braket{v^i (\eta,\vec{x}) \pi^i (\eta,\vec{y}) + \pi^i (\eta,\vec{y}) v^i (\eta,\vec{x})} = \int \frac{d^3k}{(2\pi)^3} \frac{1}{k|\eta|} e^{-i \vec{k} (\vec{x}-\vec{y})} \,. \\
	\end{aligned}
	\end{equation}
\end{itemize}

These integrals are taken over momenta that are affected by inflation, i.e. those that are sub-horizon when inflation starts and become super-horizon before it ends. These are momentum modes that satisfy:
\begin{equation}
    -\eta_0 > k^{-1} > -\eta_{\rm end}\,.
\end{equation}

When inflation ends at $\eta = \eta_{\rm end}$, mode functions are matched at the beginning of the radiation era at $\eta = -\eta_{\rm end}$. The radiation era satisfies $z''/z = 0$ and so solutions to the equation of motion (\ref{eq:vz}) are plane waves. Once the boundary conditions are imposed and taking into account that $c_S^2 = 1/3$ during the radiation era we get the solution:
\begin{equation}
\begin{aligned}
	v^r_k (\eta) = & \frac{e^{ik\eta_{\rm end}}}{\sqrt{2}} \cdot \\
	& \cdot \bigg[\frac{(-\sqrt{3} + (1+\sqrt{3}) \eta_{\rm end}  k (\eta_{\rm end}  k+i))}{2 \eta_{\rm end} ^2 k^{5/2}} e^{\frac{i}{\sqrt{3}}k(\eta+\eta_{\rm end})}\\ \label{eq:vr}
	& + \frac{(\sqrt{3} + (1-\sqrt{3}) \eta_{\rm end}  k (\eta_{\rm end}  k+i))}{2 \eta_{\rm end} ^2 k^{5/2}}e^{-\frac{i}{\sqrt{3}} k(\eta+\eta_{\rm end})}\bigg]\,.
\end{aligned}
\end{equation}

Note that this mode function is a linear combination of oscillating functions, unlike in the Minkowski vacuum, in which the oscillation affects only a global phase of the mode function. Modes significantly affected by inflation satisfy $k\eta_{\rm end} \ll 1$ so it is enough to keep leading terms in inverse powers of $(k\eta_{\rm end})$.
\begin{equation}
v_k^r (\eta) \simeq  \frac{e^{ik \eta_{\rm end}}}{i\sqrt2\,\eta_{\rm end}^2 k^{5/2}}
\sin\left(\frac{k (\eta + \eta_{\rm end})}{\sqrt3}\right)\,.
\end{equation}

The dependence on the $\sin$ function of the mode function is a general result for modes significantly affected by inflation~\cite{Mukhanov:1990me}. From (\ref{eq:vr}) one can also get the explicit form of the canonical momentum mode function $\pi^r_k(\eta)$ by using the definition $(\ref{eq:H})$.

We will consider only super-horizon modes, i.e. those that satisfy $k\eta \ll 1$ at a given time during the radiation era. For those the correlation functions are:

\begin{itemize}
    \item 1-point correlation functions
    \begin{equation}
        \braket{v^r(x,\eta)} = \braket{\pi^r(x,\eta)} = 0\,.
    \end{equation}
    \item 2-point correlation functions
    \begin{equation}\label{eq:vvr}
    \begin{aligned}
        &\braket{v^r (\eta,\vec{x}) v^r (\eta,\vec{y})} \simeq
         \int \frac{d^3k}{(2\pi)^3} e^{i\vec{k}(\vec{x}-\vec{y})}\\
        & \quad \bigg(\frac{1}{2k} + \frac{1}{2 k^3 \eta_{\rm end}^2}
         + \frac{1}{2} \left(\frac{\eta + \eta_{\rm end}}{|\eta_{\rm end}|}\right)^2 \frac{1}{k^3 \eta_{\rm end}^2} + ... \bigg)\,,
    \end{aligned}
    \end{equation}
    \begin{equation}\label{eq:ppr}
    \begin{aligned}
        &\braket{\pi^r (\eta,\vec{x}) \pi^r (\eta,\vec{y})} \simeq \int \frac{d^3 k}{(2\pi)^3} e^{i\vec{k}(\vec{x}-\vec{y})}  \\
        &\quad \bigg(\frac{k}{2} + \frac{2}{3} \left(\frac{\eta + \eta_{\rm end}}{|\eta_{\rm end}|}\right) \frac{1}{k \eta_{\rm end}^2} + ... \bigg)\,, 
    \end{aligned}
    \end{equation}
    \begin{equation}\label{eq:vpr}
    \begin{aligned}
        &\braket{v^i (\eta,\vec{x}) \pi^i (\eta,\vec{y}) + \pi^i (\eta,\vec{y}) v^i (\eta,\vec{x})} \simeq \int \frac{d^3k}{(2\pi)^3} e^{i\vec{k}(\vec{x}-\vec{y})} \\
        & \quad \bigg( \frac{1}{k \eta_{\rm end}} - \frac{1}{3} \left(\frac{\eta + \eta_{\rm end}}{|\eta_{\rm end}|}\right) \frac{1}{k^3 \eta_{\rm end}^3} + ... \bigg) \,. \\
    \end{aligned}
    \end{equation}
\end{itemize}

The quadratic term in eq. (\ref{eq:vvr}) is clearly dominant, while the expansion in eqs. (\ref{eq:ppr}) and (\ref{eq:vpr}) is a bit more involved and only the first term is shown for illustrative purposes. Hence, after inflation ends and the radiation era starts, 2-point correlation functions continue growing with conformal time $\eta$. This is a general result that can be understood as well in the squeezing formalism.

\subsection{General correlation functions in the squeezing formalism}

For general inflationary models, one can treat the time-evolution of the Bunch-Davies vacuum using the squeezing formalism, which of course can be applied to the quasi de-Sitter case as well. Such a state is characterized by the following correlation functions involving the field $v$ and its canonical conjugate $\pi$ \cite{Brandenberger:1992jh}

\begin{itemize}
	\item 1-point correlation functions
	\begin{equation}
		\braket{v(\eta, \vec{x})} = \braket{\pi(\eta, \vec{x})} = 0\, .
	\end{equation}

\item 2-point correlation functions
	\begin{equation}
	\begin{aligned}
		\braket{v(\eta,\vec{x})v(\eta,\vec{y})} &=   \int \frac{d^3k}{(2\pi)^3} e^{i\vec{k}(\vec{x}-\vec{y})} \bigg( \frac{1}{2k}\big(1+2\sinh ^2 \tau_k \\
		& -\sinh 2\tau_k \cos 2\delta _k \big) \bigg) \, , \\
		\braket{\pi(\eta,\vec{x})\pi(\eta,\vec{y})} &=  \int \frac{d^3k}{(2\pi)^3} e^{i\vec{k}(\vec{x}-\vec{y})} \bigg( \frac{k}{2}\big(1+2\sinh^2 \tau_k \\
		& +\sinh 2\tau_k \cos 2\delta_k \big) \bigg) \, ,\\ 
	\end{aligned}
	\end{equation}
	\begin{equation}
	\begin{aligned}
		&\braket{v(\eta,\vec{x})\pi(\eta,\vec{y})} =\\
		&\quad \int \frac{d^3k}{(2\pi)^3} e^{i\vec{k}(\vec{x}-\vec{y})}  \bigg(\frac{i}{2}\big(1 + i\sinh 2 \tau_k \sin 2\delta_k \big) \bigg) \, , \\[2mm]
		&\braket{\pi(\eta,\vec{y})v(\eta,\vec{x})} =\\
		&\quad \int \frac{d^3k}{(2\pi)^3} e^{i\vec{k}(\vec{x}-\vec{y})}  \bigg(\frac{-i}{2}\big(1-i\sinh 2 \tau_k \sin 2\delta_k \big) \bigg)\, .\\
	\end{aligned}
	\end{equation}
\end{itemize}

The squeezeng parameter $\tau_k$ and phase $\delta_k$ can be derived from the inflationary dynamics and the subsequent evolution in the radiation era and have a momentum-dependent expression. However, we will perform the following approximation: we will assume a random character of the phases $\delta_k$ so that integrals over $\sin 2\delta_k$ or $\cos 2\delta_k$ vanish. This is a standard procedure in the study of primordial perturbations and is justified by the effect of small self-interactions or interactions with other fields \cite{Brandenberger:1992jh, Brahma:2020zpk}. It can be seen as a coarse-graining or decoherence procedure, where the off-diagonal elements of the density matix in momentum space $\rho(\vec{k},-\vec{k},\vec{p},\vec{-p})$ decay. Nevertheless, our results would not change significantly if the random phase approximation was not performed. We leave this discussion in the Appendix for the interested reader.


After averaging over the phases the correlators become:
\begin{equation}
\begin{aligned}
	& \braket{v(\eta,\vec{x})v(\eta,\vec{y})} = \int \frac{d^3k}{(2\pi)^3} e^{i\vec{k}(\vec{x}-\vec{y})} \frac{1}{2\omega_k}\left(1+2\sinh ^2 \tau_k \right)\, ,\\
	& \braket{\pi(\eta,\vec{x})\pi(\eta,\vec{y})} = \int \frac{d^3k}{(2\pi)^3} e^{i\vec{k}(\vec{x}-\vec{y})} \frac{\omega_k}{2}\left(1+2\sinh^2 \tau_k \right)\, ,\\[2mm]
	& \braket{v(\eta,\vec{x})\pi(\eta,\vec{y}) + \pi(\eta,\vec{y})v(\eta,\vec{x})} = 0 \, .
\end{aligned}
\end{equation}

Because of the term $\left(1+2\sinh^2 \tau_k \right)$ we get an effective enhancement of the field and conjugate correlations for those momentum modes that are affected by inflation, i.e. those that satisfy
\begin{equation}
	-\eta_0 > k^{-1} > -\eta_{\rm end}\,,
\end{equation}

The affected modes are thus those with wavelength smaller than the horizon when inflation starts and larger than the horizon when it ends.

One could ask what should be the correlators for modes that are not squeezed. It is clear that, for modes with small wavelength $k^{-1} < -\eta_{\rm end}$, we can take them to be equal to those of the Minkowski vacuum due to the Bunch-Davies prescription. However, there is little if anything we can say about those modes with large wavelength $k^{-1} > -\eta_0$ as they were already super-horizon when inflation started. Those modes should have physical effects only at extremely large scales, much larger than the observable universe. We expect them to give an irrelevant contribution to the correlator and thus we will treat them as if they were in the Minkowski vacuum as well.

In our discussion we will not pay too much attention to the particular inflationary dynamics. Instead, we will take the following quite general result for the squeezing parameter during inflation \cite{Albrecht:1992kf}:
\begin{equation}
	\tau^i_k = \log\left(\frac{1}{-\eta k}\right) \quad
	\textrm{for} \quad
	-\eta_0 > k^{-1} > -\eta\,,
\end{equation}

and $\tau_k = 0$ otherwise. Notice that once inflation ends, this squeezing parameter will have a dependence on the conformal time at the end of inflation, but not at its beginning. Furthermore, during the radiation era the quantum state undergoes additional squeezing, so that its parameter is given by \cite{Albrecht:1992kf}:

\begin{equation}
    \tau^r_k = \log\left(\frac{1}{|\eta_{\rm end}| k}\right) + \log \left( \frac{\eta}{|\eta_{\rm end}|} \right) 
    \quad \textrm{for} \quad k^{-1} > \eta \,.
\end{equation}

This additional term stops growing once the mode re-enters the horizon at $\eta = k^{-1}$ and reaches $\tau^r_k = 2\log(|\eta_{\rm end}|k)$. In the rest of the paper we will restrict ourselves to modes that remain super-horizon. It is important to notice that modes that are sub-horizon when inflation ends are not squeezed during the radiation era.

The enhancement of the correlation functions for modes affected by inflation is then
\begin{equation}
	1+ 2 \sinh^2 \tau_k = \frac{1}{2} \left(\frac{1}{k^2 \eta_{\rm end}^2} \left( \frac{\eta}{\eta_{\text{end}}} \right)^2 +k^2\eta_{\rm end}^2 \left( \frac{\eta_{\text{end}}}{\eta} \right)^2 \right) \, .
\end{equation}

The second term can be neglected because of the condition $k\eta_{\text{end}} \ll 1$ and the fact that $\eta > |\eta_{\text{end}}|$. The correlators in the random phase approximation for modes that are affect by inflation and stay super-horizon at conformal time $\eta$ in the radiation era are then given by:
\begin{equation}
\begin{aligned}
	\braket{v(x),v(y)} = &\int_k \frac{d^3k}{(2\pi)^3} \frac{1}{4k} \frac{1}{k^2 \eta_{\rm end}^2} \left( \frac{\eta}{\eta_{\text{end}}} \right)^2 e^{i\vec{k}\cdot(\vec{x}-\vec{y})} \, ,\\
	\braket{\pi(x),\pi(y)} = &\int_k \frac{d^3k}{(2\pi)^3} \frac{k}{4} \frac{1}{k^2 \eta_{\rm end}^2}\left( \frac{\eta}{\eta_{\text{end}}} \right)^2 e^{i\vec{k}\cdot(\vec{x}-\vec{y})}\, .
\end{aligned}
\end{equation}

\begin{figure}
    \centering
    \includegraphics[width=\linewidth]{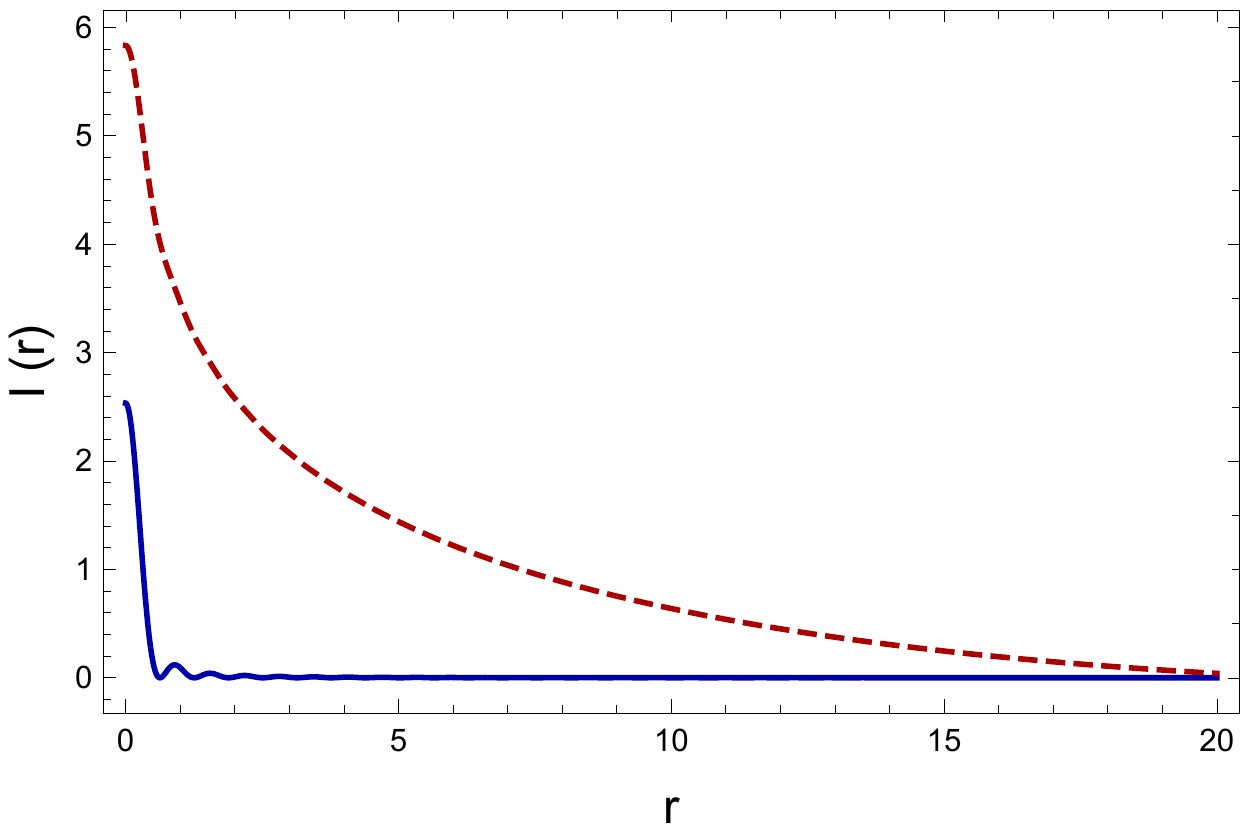}
    \caption{An example of the difference between the enhanced correlator (red dashed line) and the Minkowski one (restricted to inflationary modes, blue line) for $-\eta_0 = 10$, $-\eta_{\rm end}= 0.1$, $\eta = \eta_{\text{end}}$. The Minkowski correlator decays very fast for distances larger than the scale of the largest momentum, while the enhanced correlator has a much slower decay.}
    \label{fig:Correlator}
\end{figure}

The enhancement of the 2-point correlation functions is translated into a slower decay. The long range behavior of the Minkowski correlation functions is known to be \cite{Bombelli:1986rw,Shiba:2010dy}
\begin{equation}
	\int \frac{d^3k}{(2\pi)^3}\frac{1}{k}e^{i\vec{k}\cdot\vec{r}} \sim r^{-2} \quad \textrm{and} \quad \int \frac{d^3k}{(2\pi)^3}k e^{i\vec{k}\cdot\vec{r}} \sim r^{-4}\,,
\end{equation}
where $r = |\vec{x}-\vec{y}|$. The result is similar when considering other powers of $k$ in the integrand
\begin{equation}
	\int \frac{d^3k}{(2\pi)^3} k^{\alpha} e^{i\vec{k}\cdot\vec{r}} = r^{-(3+\alpha)} \quad
	\textrm{for} \quad 
	\alpha > -3\,,
\end{equation}
and thus correlations decay fast with distance. This is also true for several of the enhanced terms, as they satisfy this form with $\alpha > -3$. However, there is one term in the field-field correlation function that has $\alpha = -3$, namely
\begin{equation}
	I(r) = \frac{1}{4\eta_{\rm end}^2} \left( \frac{\eta}{\eta_{\text{end}}} \right)^2 \int_{k\in {\rm inf}} \frac{d^3 k}{(2\pi)^3} \frac{1}{k^3} e^{i\vec{k}\cdot\vec{r}}\,.
\end{equation}

In the long-range regime, this integral has an analytic expression
\begin{equation}
\begin{aligned}
	I(r) = & \frac{1}{8\pi^2 \eta_{\rm end}^2} \left( \frac{\eta}{\eta_{\text{end}}} \right)^2 \bigg[-\text{Ci}\left(\frac{r}{-\eta_0}\right)+ \text{Ci}\left(\frac{r}{\eta}\right)\\
	& +\frac{-\eta_0}{r} \sin \left(\frac{r}{-\eta_0}\right)-\frac{\eta }{r} \sin \left(\frac{r}{\eta}\right) \bigg]\,,
	\label{eq:exact}
\end{aligned}
\end{equation}
where Ci is the cosine integral defined as
\begin{equation}
	\textrm{Ci} (x) = -\int_x^{\infty}\frac{\cos t dt}{t} = \gamma + \log x + \int_0^x \frac{\cos t -1}{t} dt \,.
\end{equation}

And $\gamma = 0.577216...$ is the Euler-Mascheroni constant. Because of the logarithmic behavior of the cosine integral, this term of the field-field correlator decays logarithmically with distance until $r \simeq -\eta_0$, i.e. the enhancement happens only up to length-scales comparable to the wavelength of the longest momentum modes affected by inflation.

If inflation lasts for a finite number of e-folds the correlation vanishes at infinity
\begin{equation}
	\lim_{r\rightarrow \infty} I(r) = 0\,.
\end{equation}

The expression above is not very intuitive, but we can approximate it by assuming that $ r \ll -\eta_0$  which is a reasonable approximation until distances reach the scale of the horizon at the beginning of inflation. Then we have
\begin{equation}
\begin{aligned}
	& \text{Ci}\left(\frac{r}{-\eta_0} \right) \simeq \gamma + \log\left(\frac{r}{-\eta_0}\right) \,, \\
	& \sin \left(\frac{r}{-\eta_0} \right) \simeq \frac{r}{-\eta_0} \,.
\end{aligned}
\end{equation}
Then
\begin{equation}\label{eq:approx}
\begin{aligned}
	I(r) \simeq & \frac{1}{8\pi^2 \eta_{\rm end}^2} \left( \frac{\eta}{\eta_{\text{end}}} \right)^2 \bigg[ \text{Ci} \left(\frac{r}{\eta}\right) - \gamma + \log\left(\frac{-\eta_0}{r}\right)\\
	& -\left(\frac{\eta}{r} \right) \sin\left( \frac{r}{-\eta_{\rm end}} \right) + 1 \bigg] \,.
\end{aligned}
\end{equation}

Since we are limiting ourselves to super-horizon scales, we can also assume $r \gg \eta$ and perform further approximations
\begin{equation}
\begin{aligned}
 & \text{Ci} \left( \frac{r}{\eta} \right) \simeq 0 \,, \\
 & \left(\frac{\eta}{r} \right) \sin\left( \frac{r}{\eta} \right) \simeq 0 \,. \\
\end{aligned}
\end{equation}
And we get the expression 
\begin{equation}\label{eq:approx2}
	I(r) \simeq \frac{1}{8\pi^2 \eta_{\rm end}^2} \left( \frac{\eta}{\eta_{\text{end}}} \right)^2 \left[  \log\left(\frac{-\eta_0}{r}\right)  + 1 - \gamma \right]\,.
\end{equation}

\begin{figure}
   \includegraphics[width=\linewidth]{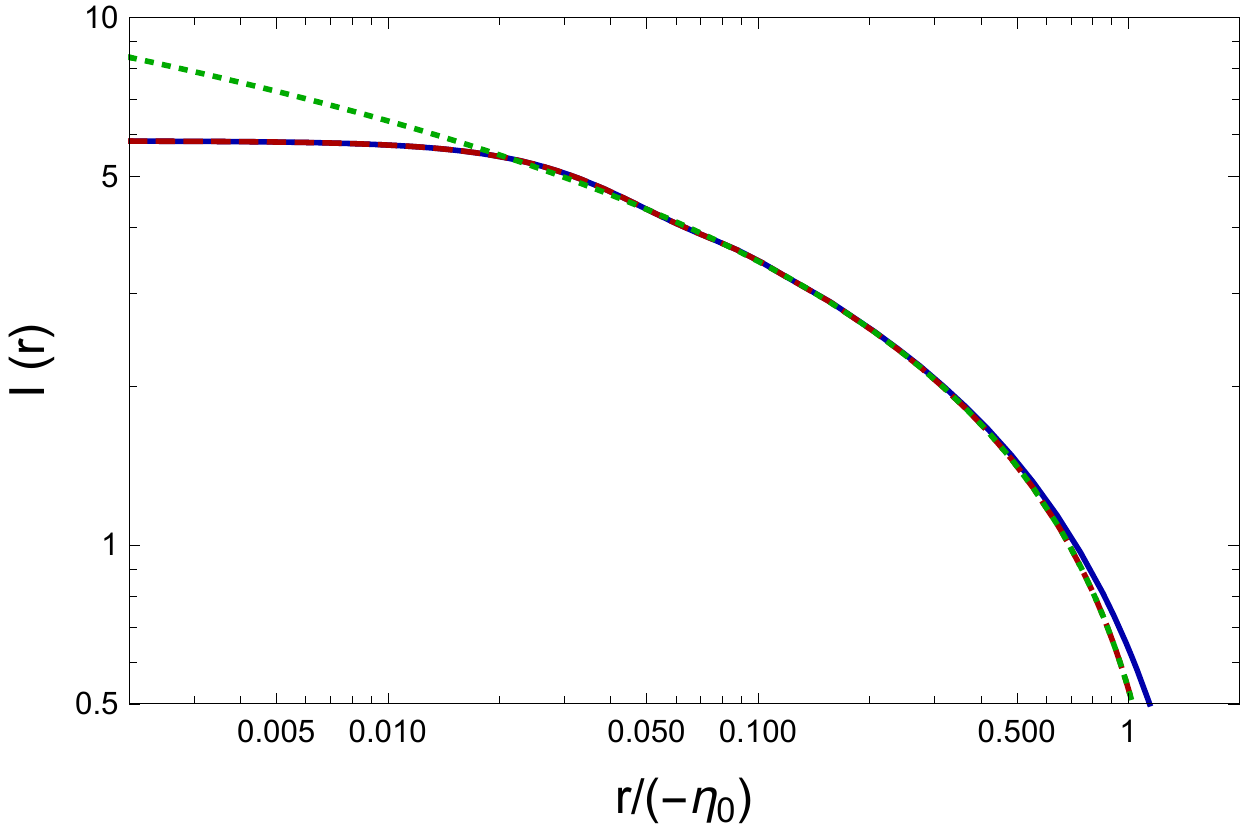}
    \caption{Comparison between the exact expression for the enhanced correlator $I(r)$ Eq.~(\ref{eq:exact}) (blue line), its approximation Eq.~(\ref{eq:approx}) (red dashed line), and the further logarithmic approximation Eq.~(\ref{eq:approx2}) (green dotted line), for $-\eta_0 = 10, -\eta_{\rm end} = 0.1$, $\eta = \eta_{\rm end}$. The agreement is excellent until distances of the order of $r/(-\eta_0)\sim1$, where both approximations start to slowly diverge.}
    \label{fig:approx}
\end{figure}

Physically, field correlations are enhanced in those momentum modes affected by inflation. This can be understood as modes being stretched out from small scales and then occupied due to particle creation. The next step will be to review the connection between correlation and entropy or information.


\section{IV. Entropy of the scalar field}

Let us now revisit the problem of computing the entropy of a spatial region for a scalar field in a gaussian state. Gaussian states are simple enough for a systematic method to be developed but already include important states such as the Minkowski vacuum or the squeezed state from inflation, which we are considering in this paper. In order to do so, we will take advantage of the fact that Gaussian states can be fully characterized by its equal-time 1-point and 2-point correlation functions. We refer the reader to \cite{Casini:2007bt,Casini:2009sr,Berges:2017hne} for additional details.

The computation of the entropy becomes particularly simple in the case of vanishing expected values
\begin{equation}
	\braket{v(\vec{x})} = 0 \quad \braket{\pi(\vec{x})} = 0\,.
\end{equation}
And vanishing symmetrized 2-point cross-correlation function
\begin{equation}
	\braket{v(\vec{x})\pi(\vec{y}) + \pi(\vec{y})v(\vec{x})} = 0\,.
\end{equation}
This is the case for the squeezed state of the curvature perturbation field once the averaging over phases is performed. The other 2-point correlation functions are given by the operator kernels
\begin{equation}
	X(\vec{x},\vec{y}) = \braket{v(\vec{x})v(\vec{y})} \quad P(\vec{x},\vec{y}) = \braket{\pi(\vec{x})\pi(\vec{y})}\,.
\end{equation}
Then one defines the operator
\begin{equation}
	\Lambda_\Omega = X\cdot P\,,
\end{equation}
where the operator product is equivalent to a convolution of the kernels
\begin{equation}
\Lambda_{\Omega} (\vec{x},\vec{y}) = \int_{\Omega^C} d^3 z X(\vec{x},\vec{z}) P(\vec{z},\vec{y})\,,
\end{equation}
where the region $\Omega^C$ comprises the local d.o.f. that we wish to trace out, thereby being left with an operator kernel defined on $\Omega$ only. Then the  entropy of the complementary region $\Omega$ can be computed as
\begin{equation}
\begin{aligned}
	S_{\Omega} = & \textrm{Tr} \bigg[ \left(\sqrt{\Lambda_{\Omega}}+1/2)\right) \log \left(\sqrt{\Lambda_{\Omega}}+1/2)\right) \\
	& - \left(\sqrt{\Lambda_{\Omega}}-1/2)\right) \log \left(\sqrt{\Lambda_{\Omega}}-1/2)\right) \bigg] \,. \\
\end{aligned}
\end{equation}

Note that the kernel of the square root is not the square root of the kernel and so we cannot give a closed expression for the kernel $\sqrt{\Lambda_{\Omega}} (\vec{x},\vec{y})$. However, in order to compute numerically this complicated expression, we do not need to know it. Instead, one needs to solve the eigenvalue problem for $\Lambda_{\Omega}$, i.e. find those  $\lambda_i$ for which
\begin{equation}
	\int_{\Omega} d^3y \Lambda_{\Omega} (\vec{x},\vec{y}) f_i(\vec{x}) = \lambda_i f_i(\vec{y})\,,
\end{equation}
where $f_i$ is the eigenfunction of $\Lambda_{\Omega}$ with eigenvalue $\lambda_i$. Then one has
\begin{equation}
\begin{aligned}
    S_{\Omega} = & \sum_i h(\lambda_i) =\\
     = & \sum_i \bigg[ \left(\sqrt{\lambda_i}+1/2)\right) \log \left(\sqrt{\lambda_i}+1/2)\right)\\
	& - \left(\sqrt{\lambda_i}-1/2)\right) \log \left(\sqrt{\lambda_i}-1/2)\right) \bigg]\,.
\end{aligned}
\end{equation}

Nevertheless, we will compute the mutual information perturbatively, without needing to obtain exact results for $S_{\Omega}$. That is, if we take $\Omega = A \cup B$, where $A$ and $B$ are, then we have that
\begin{equation}
	S_{A\cup B} (r) = S_A + S_B - I(A,B)(r)\,.
\end{equation}

This method was introduced by Noburo Shiba in  Ref.~\cite{Shiba:2010dy}. We will adapt it to the case of cosmological perturbations by using a more general formalism (valid for arbitrary Gaussian states, not only the vacuum) and applying it to the quantum state following inflation. One expects that the mutual information should vanish at infinite distance
\begin{equation}
	\lim_{r \rightarrow \infty} I(A,B)(r) = 0 \,.
\end{equation}
And conversely
\begin{equation}
	\lim_{r\rightarrow \infty} S_{A\cup B} (r) = S_A + S_B \,.
\end{equation}

The idea then is to expand perturbatively the joint entropy $S_{A \cup B}$ as the individual entropies $S_A$ and $S_B$ and a term involving functions of the distance that vanish at infinity. This can be already done at the operator level by identifying what terms in $\Lambda_{A\cup B}$ depend on the distance $r$ and expanding them.

For the case at hand, $\Lambda$ will carry both contributions from the Minkowski vacuum as well as the squeezed modes. The former will be responsible for a mutual information that scales as $r^{-4}$ and thus is of no interest to us. The latter, however, will be responsible for an enhanced mutual information that decays logarithmically.

\section{V. The perturbative computation}

We are interested in perturbative solutions to the eigenvalue problem
\begin{equation}
	\int_{\Omega} d^3y \Lambda_{\Omega} (\vec{x},\vec{y}) f_i(\vec{x}) = \lambda_i f_i(\vec{y})\,,
\end{equation}
with the choice
\begin{equation}
    \Omega = A \cup B\,,
\end{equation}
where $A$ and $B$ are two disjoint regions of size $R_A$ and $R_B$ separated by a large distance $r$ such that $r \gg R_A,R_B$. Both regions need not be spherical, although this is the simplest and perhaps most interesting application.

We will find these perturbative solutions by following the next steps
\begin{itemize}
	\item We identify the perturbative and non-perturbative contributions.
	\item We identify the leading perturbative contribution. In our case this will mean keeping only the enhancement of the correlation functions.
\end{itemize}

The behavior of $\Lambda_{\Omega}$ depends on whether $x$ and $y$ belong to the regions $A$ or $B$. We represent this in matrix form
\begin{equation}
	\Lambda_{\Omega}(x,y) =
	\begin{pmatrix}
		\Lambda_{\Omega}(\vec{x}_a, \vec{y}_a) & \Lambda_{\Omega}(\vec{x}_a,\vec{y}_b)\\
		\Lambda_{\Omega}(\vec{x}_b, \vec{y}_a) & \Lambda_{\Omega}(\vec{x}_b,\vec{y}_b)\\
	\end{pmatrix}\,.
\end{equation}

It is understood that $\vec{x}_a, \vec{y}_a \in A$ and $\vec{x}_b, \vec{y}_b \in B$.

\subsection{Perturbative part}

We take first a look at the off-diagonal terms, as they clearly involve points belonging to different regions. First, we rewrite the off-diagonal terms using the relation
\begin{equation}
	\Lambda_{\emptyset}(\vec{x}_{a/b},\vec{y}_{b/a}) = \delta^{(3)}(\vec{x}_{a/b},\vec{y}_{b/a}) = 0\,,
\end{equation}

where $\emptyset = \{ \mathbb{R}^3 \}^C$ is the empty set. We will use the notation ${a/b}$ to mean "$a$ or $b$" and the order will matter if it appears several times in an equation. Then the Dirac delta equals $0$ because $\vec{x}_{a/b} \neq \vec{y}_{b/a}$ when one point belongs to $A$ and the other belongs to $B$. We can then rewrite
\begin{equation}
	\Lambda_{\Omega}(\vec{x}_{a/b},\vec{y}_{b/a}) =-\Lambda_{\Omega^C} (\vec{x}_{a/b},\vec{y}_{b/a})\,,
\end{equation}
with
\begin{equation}
\begin{aligned}
	\Lambda_{\Omega^C} (\vec{x}_{a/b},\vec{y}_{b/a}) = & \int_A d^3z_a X(\vec{x}_{a/b},\vec{z}_a) P(\vec{z}_a,\vec{y}_{b/a})\\
	& + \int_B d^3z_b X(\vec{x}_{a/b},\vec{z}_b) P(\vec{z}_b,\vec{y}_{b/a})\,.
\end{aligned}
\end{equation}

Strictly speaking, $\Lambda_{\emptyset} \simeq \delta $ but the equality is not exact. The difference is small from the operator point of view and we will neglect it. It is also an artifact of assuming random phases.

Notice that for each of the integrals, either the kernel $X(x,y)$ or $P(x,y)$ has a long-distance behavior, i.e. it is evaluated at points belonging to different regions. Both kernels have the form of a Fourier transform, regardless of whether we consider the Minkowksi or the squeezed correlators
\begin{equation}
\begin{aligned}
	&X(\vec{x},\vec{y}) = \int \frac{d^3k}{(2\pi)^3} X(k) e^{i\vec{k}(\vec{x}-\vec{y})} \,, \\
	&P(\vec{x},\vec{y}) = \int \frac{d^3k}{(2\pi)^3} P(k) e^{-i\vec{k}(\vec{x}-\vec{y})}\,, 
\end{aligned}
\end{equation}
where the only dependence on the direction of $\vec{k}$ is encoded in the exponential. In the long-distance regime we can approximate
\begin{equation}
\begin{aligned}
	\int d\theta \sin \theta e^{ik|\vec{x}-\vec{z}|\cos\theta}\simeq \int d\theta \sin \theta e^{ikr\cos\theta} = \frac{2\sin(kr)}{kr}
\end{aligned}
\end{equation}
and the integral over $z$ will be irrelevant for this kernel since
\begin{equation}
	|\vec{a}-\vec{b}| \simeq r \quad \textrm{for} \quad \forall \vec{a}\in A,\vec{b}\in B \,.
\end{equation}

Hence, we will approximate from now on
\begin{equation}
	X(\vec{x}_{a/b},\vec{y}_{b/a}) \simeq I(r) \,.
\end{equation}
and we will keep only terms involving $I(r)$ in the off-diagonal components of $\Lambda_{\Omega}$, since they are the leading perturbative contribution. This leaves us with
\begin{equation}
\begin{aligned}
	&\delta\Lambda_{\Omega} (r) = - I(r) \begin{pmatrix}
		0 & \int_B d^3 z_b P(\vec{z}_b,\vec{y}_b)\\
		\int_A d^3 z_a P(\vec{z}_a,\vec{y}_a) & 0\\
	\end{pmatrix}\,.
\end{aligned}
\end{equation}

\subsection{Non-perturbative part}

The non-perturbative part of $\Lambda_{\Omega}$ needs some refinement. One would think first to simply choose its block-diagonal components
\begin{equation}
	\Lambda_{\Omega}^D = \Lambda_{\Omega}(\vec{x}_{a/b},\vec{y}_{a/b}) \,.
\end{equation}
But this still depends on $r$, as it integrates over $z\in \Omega^C = (A \cup B)^C$. Instead, we define the non-perturbative part as the limit
\begin{equation}
\begin{aligned}
	\Lambda_{\Omega}^0 & = \lim_{r\rightarrow \infty} \Lambda_{\Omega}^D =\\
	& = \begin{pmatrix}
		\int_{A^C} d^3z X(\vec{x}_a,\vec{z})P(\vec{z},\vec{y}_a) & 0\\
		0 &  \int_{B^C} d^3z X(\vec{x}_b,\vec{z})P(\vec{z},\vec{y}_b)\nonumber
	\end{pmatrix}\,.
\end{aligned}
\end{equation}

The difference is given by a perturbative contribution that decays faster than $I(r)$, as it decays at most as slow as $I(r)$ times an additional perturbative term
\begin{equation}
\begin{aligned}
	&\Lambda_{\Omega}^0 - \Lambda_{\Omega}^D = \\
	&= \begin{pmatrix}
		\int_{B} d^3z_b X(\vec{x}_a,\vec{z}_b)P(\vec{z}_b,\vec{y}_a) & 0\\
		0 &  \int_{A} d^3z X(\vec{x}_b,\vec{z}_a)P(\vec{z}_a,\vec{y}_b) \nonumber
	\end{pmatrix}\,.
\end{aligned}
\end{equation}

Since $X$ will decay at most as slow as $I(r)$ and $P$ will decay as some inverse power of $r$, it is clear that $\Lambda_{\Omega}^0 - \Lambda_{\Omega}^D$ is a negligible perturbative term.

We have now a well-posed perturbative approach for the eigenvalue problem.

\subsection{Non-hermitian perturbation theory}

The first thing we should notice when taking the perturbative approach is that neither $\Lambda_{\Omega}^0$ nor $\delta \Lambda_{\Omega}$ are symmetric operators. This means that, in principle, it is not guaranteed that $\Lambda_{\Omega}$ is diagonalizable or that the computation of its eigenvalues admits the usual perturbative treatment. In practice, one can argue that $\Lambda_{\Omega}$ is diagonalizable \cite{Berges:2017hne}, nevertheless the issue of applying perturbation theory remains. For a detailed treatment of non-hermitian perturbation theory we refer the reader to \cite{PhysRevC.6.114}. We will need to work with symmetrized forms of both operators, which we will achieve by introducing the following operator
\begin{equation}
	P_0 = \lim_{r\rightarrow \infty} P  =\begin{pmatrix}
		P(\vec{x}_a,\vec{y}_a) & 0 \\
		0 & P(\vec{x}_b,\vec{y}_b)
	\end{pmatrix}\,,
\end{equation}
so that the operator $P_0\Lambda_{\Omega}$ is indeed symmetric. Let us see why. For the perturbative part it is pretty straightforward:

\begin{equation}
\begin{aligned}
	&P_0 \delta \Lambda_{\Omega} (\vec{x}_a,\vec{y}_b) =\\
	& = - \int_A d^3z_a \int_B d^3z_b P(\vec{x}_a,\vec{z}_a)X(\vec{z}_a,\vec{z}_b)P(\vec{z}_b,\vec{y}_b)\\
	& \simeq - I(r) \int_A d^3z_a \int_B d^3z_b P(\vec{x}_a,\vec{z}_a) P(\vec{z}_b,\vec{y}_b) \,.
\end{aligned}
\end{equation}

And:

\begin{equation}
\begin{aligned}
	& P_0 \delta \Lambda_{\Omega} (\vec{x}_b,\vec{y}_a)\\
	& = - \int_A d^3z_a \int_B d^3z_b P(\vec{x}_b,\vec{z}_b)X(\vec{z}_b,\vec{z}_a)P(\vec{z}_a,\vec{y}_a)\\
	& \simeq  - I(r) \int_A d^3z_a \int_B d^3z_b P(\vec{x}_b,\vec{z}_b)P(\vec{z}_a,\vec{y}_a)\,.
\end{aligned}
\end{equation}

It is clearly symmetric since $P$ is symmetric. The non-perturbative part is perhaps less obvious:

\begin{equation}
\begin{aligned}
	P_0 \Lambda_{\Omega}^0 (\vec{x}_a,\vec{y}_a) = \int_A d^3z_a \int_{A^C}d^3z P(\vec{x}_a,\vec{z}_a)X(\vec{z}_a,\vec{z})P(\vec{z},\vec{y}_a)\,.
\end{aligned}
\end{equation}

And

\begin{equation}
\begin{aligned}
	P_0 \Lambda_{\Omega}^0 (\vec{x}_b,\vec{y}_b) = \int_B d^3z_b \int_{B^C}d^3z P(\vec{x}_b,\vec{z}_b)X(\vec{z}_b,\vec{z})P(\vec{z},\vec{y}_b)\,.
\end{aligned}
\end{equation}

But we can now make use of the relation that was argued previously:

\begin{equation}
	\int_{\mathbb{R}^3} d^3z X(\vec{x},\vec{z})P(\vec{z},\vec{y}) = \delta^{(3)}(\vec{x}-\vec{y})\,.
\end{equation}

Then we can rewrite the non-perturbative part as:

\begin{equation}
\begin{aligned}
	& P_0 \Lambda_{\Omega}^0 (\vec{x}_a,\vec{y}_a)= \\
	& -\int_A d^3z_1 \int_{A}d^3z_2 P(\vec{x}_a,\vec{z}_1)X(\vec{z}_1,\vec{z}_2)P(\vec{z}_2,\vec{y}_a)\\
		& + P(\vec{x}_a,\vec{y}_a) \,.
\end{aligned}
\end{equation}

And analogously for $P(\vec{x}_b,\vec{y}_b)$ one checks that the result is symmetric.

Next, let us discuss the eigenvalue problem for $\Lambda_{\Omega}^0$ first. We know that it is diagonalizable and has real eigenvalues \cite{Berges:2017hne}, but it is still a non-symmetric operator. Therefore, its right and left eigenvectors do not need to coincide.  Let us consider a right eigenvector $f_i$

\begin{equation}
	\Lambda^0_{\Omega} f^0_i = \lambda_i f^0_i \,.
\end{equation}

We can apply $P_0$ on the left and define a new set of vectors $\tilde{f}^0_i := P_0 f_i^0$. Notice what happens if we compute:

\begin{equation}
	P_0 \Lambda^0_{\Omega} f^0_i = \lambda^0_i P_0 f^0_i \equiv \lambda^0_i \tilde{f}^0_i \,.
\end{equation}

It turns out that $\tilde{f}_i^0$ are left eigenvectors of $\Lambda^0_{\Omega}$

\begin{equation}
	\lambda^0_i \tilde{f}^0_i = P_0 \Lambda_{\Omega}^0 f^0_i =  \Lambda_{\Omega}^{0\dagger} P_0 f^0_i = \Lambda_{\Omega}^{0 \dagger} \tilde{f}^0_i \,.
\end{equation}

For the perturbation theory to work, we would like this set of left and right eigenvectors to form a complete biorthonormal set, i.e. that the following identity is satisfied:

\begin{equation}
	\tilde{f}^{0 \dagger}_i f^0_j = f^{0 \dagger}_i P_0 f^0_j = \delta_{ij} \,.
\end{equation}

Let us see when this is true, starting from the fact that $P_0\Lambda$ is a symmetric operator
\begin{equation}
	0 = f^{0 \dagger}_i P_0 \Lambda_{\Omega}^0 f_j - f^{0 \dagger}_i \Lambda^{0 \dagger}_{\Omega} P_0 f_j = (\lambda_j-\lambda_i) \tilde{f}^{0 \dagger}_i f^0_j\,,
\end{equation}
which means that, if the eigenvalues are non-degenerate, then the set of left and right eigenvalues is guaranteed to be biorthonormal. If they are degenerate, one has to look into it more carefully.

We have the intuitive notion from QM that degeneracy arises when a symmetry is present. The corresponding transformation allows us to add additional labels to the degenerate eigenstates and also transform between them. Under which transformations is $\Lambda_{\Omega}^0$ invariant? Let us think of the space-time symmetries, which are actually restricted to spatial symmetries, i.e. 3-dimensional rotations and translations, since we are working with equal time correlators.

Translational symmetry is clearly broken by the choice of the regions A and B. It may be only partially broken if these regions are infinite in some direction, but this is not of interest for the case at hand. Then we are left with rotational symmetry only, which is a symmetry only of $\Lambda_{\Omega}^0$ restricted to either $A$ or $B$ when these are in turn spherically symmetric regions. In additional to this restricted rotational symmetry, the permutation $A \leftrightarrow B$ is also a symmetry if $A$ and $B$ have the same size and shape and this adds an additional degeneracy.

How can we know that this degeneracy brought by symmetry transformations $T$ is not harmful? The key is that restricted rotations and permutations commute with $P_0$, which is the operator that maps between left and right eigenvectors.
\begin{equation}
	[T,P_0] = 0\,.
\end{equation}
Recall the discussion on complete sets of commuting observables in Quantum Mechanics. Here, because we are dealing with a non-hermitian operator that plays the role of a hamiltonian, not only do we need symmetry (i.e. $[\Lambda,T] = 0$) but also the commuting relation above in order to guarantee the existence of a complete biorthonormal set of eigenstates. It is clear that $P_0$ is both invariant under restricted rotations and permutations and this is why it commutes with $T$. Permutations are really not an issue, because it is clear that eigenfunctions defined on different regions $A$ and $B$ are orthogonal. Due to rotational symmetry, we can label the right eigenvectors with degenerated eigenvalue according to its angular momentum
\begin{equation}
	f_{ilm} = f_i Y_{lm}\,,
\end{equation}
where $Y_{lm}$ are the spherical harmonics. Furthermore, the left eigenvectors are
\begin{equation}
	\tilde{f}_{ilm} = P_0 f_{ilm} = P_0 f_i Y_{lm} = \tilde{f}_i Y_{lm}\,.
\end{equation}

Since $P_0$ commutes with rotations. This guarantees now the biorthonormality relation
\begin{equation}
	\tilde{f}^{0\dagger}_{ilm} f^0_{jl'm'} = \tilde{f}^{0\dagger}_i f^0_j Y^*_{lm}Y_{l'm'} = \delta_{ij} \delta_{ll'} \delta_{mm'}\,.
\end{equation} 
And therefore it is guaranteed that $\Lambda^0_{\Omega}$ is diagonalizable and has a complete biorthogonal set of eigenvectors. We will need this later and in particular we will need the resolution of identity
\begin{equation}
	\sum_{ilm} \tilde{f}^0_{ilm} f^{0 \dagger}_{ilm} = 1 \,.
\end{equation}

\subsection{Computation}

Let us now deal with the perturbation theory itself. We will keep first- and second-order perturbations to the eigenvalues
\begin{equation}
	\lambda_i = \lambda_i^0 + \delta \lambda_i^1 + \delta \lambda_i^2 \,.
\end{equation}

And then the entropy can be computed in perturbation theory
\begin{equation}
\begin{aligned}
	S_{AB} = & \sum_i h(\lambda_i) = S_A + S_B +\\
	&  +  \sum_i \left[\delta\lambda_i \frac{dh}{d\lambda_i}\bigg|_{\lambda_i=\lambda_i^0} + \frac{1}{2}(\delta\lambda_i)^2 \frac{d^2h}{d\lambda_i^2}\bigg|_{\lambda_i=\lambda_i^0}\right]\,,
\end{aligned}
\end{equation}

where $\delta\lambda_i = \delta\lambda_i^1 + \delta\lambda_i^2$ is the combined first- and second-order perturbation and we simply denote by $h$ the function of the eigenvalues that delivers the entropy. We can clearly identify the mutual information as the third term in the RHS. We will see that the first order perturbation to the entropy vanishes and so the second-order perturbation becomes the most relevant one. The following lines are to great extent a reproduction of the results from \cite{Shiba:2010dy}.

We will try to keep $R_1 \neq R_2$ during the whole computation in order to keep it as general as possible. In fact, we will keep the regions $A$ and $B$ of arbitrary shape. Recall that the non-perturbative operator $\Lambda_{\Omega}^0$ is divided in two blocks, affecting either region $A$ or  $B$. Each of these blocks may have some common and some different eigenvalues. Then, let us introduce extra indices to take this into account, as well as other possible degeneracies. We label the eigenvalues in increasing order, i.e. $\lambda^0_m > \lambda^0_n$ when $m>n$ and the right eigenvectors with eigenvalue $\lambda^0_m$ as
\begin{equation}
	f^0_{m1\alpha} = \begin{pmatrix}
		f^0_{m1\alpha}(\vec{x}_a)\\
		0\\
	\end{pmatrix} \quad 
	f^0_{m2\beta} = \begin{pmatrix}
 		0\\
		f^0_{m2\beta}(\vec{x}_b)
	\end{pmatrix}\,.
\end{equation}
being $\alpha$ and $\beta$ some possible  degeneracies. With this notation, the orthogonality property is written as
\begin{equation}
	\tilde{f}^{0 \dagger}_{mi\alpha} f^0_{nj\beta} = \delta_{mn}\delta_{ij}\delta_{\alpha \beta}\,.
\end{equation}

The right eigenvector $f_{m\gamma}$ of the full operator $\Lambda_{\Omega}$ is a linear combination of the eigenvectors of the blocks plus perturbations
\begin{equation}
\begin{aligned}
	f_{m\gamma} & = \sum_{\alpha} a_{\gamma \alpha} f^0_{m1\alpha} + \sum_{\beta} b_{\gamma \beta} f^0_{m2\beta} + f^1_{m\gamma} + f^2_{m\gamma}\\
	&\equiv \xi ^0_{m\gamma} + f^1_{m\gamma} + f^2_{m\gamma}\,.
\end{aligned}
\end{equation}

Note that if $\lambda^0_m$ is not a common eigenvalue of both blocks, then either the $a_{\gamma \alpha}$ or the $b_{\gamma \beta}$ coefficients vanish. We can now plug the perturbative expansion of the right eigenvector $f_{m\gamma}$ in the eigenvalue equation to find
\begin{equation}
	\left(\Lambda_{\Omega}^0 + \delta \Lambda_{\Omega}\right) f_{m\gamma} = \left( \lambda^0_m + \delta \lambda^1_{m\gamma} + \delta \lambda^2_{m\gamma} \right) f_{m\gamma}\,.
\end{equation}

The first order perturbation equation is obtained by neglecting second order perturbations and pluging in the solution to the unperturbed eigenvalue equation
\begin{equation}
	\Lambda_{\Omega}^0 f^1_{m\gamma} + \delta\Lambda_{\Omega} \xi^0_{m\gamma} = \lambda^0_m f^1_{m\gamma} + \delta\lambda^1_{m\gamma} \xi^0_{m\gamma}\,.
\end{equation} 

Similarly, we obtain the second order perturbation equation
\begin{equation}
	\Lambda_{\Omega}^0 f^2_{m\gamma} + \delta\Lambda_{\Omega} f^1_{m\gamma} = \lambda^0_m f^2_{m\gamma} + \delta\lambda^1_{m\gamma}f^1_{m\gamma} + \delta\lambda^2_{m\gamma} \xi^0_{m\gamma}\,.
\end{equation}

We take now the first order perturbation equation and multiply it by $\tilde{f}^{0\dagger}$ on the left
\begin{equation}
	\tilde{f}^{0\dagger}_{mj\gamma'} \Lambda_{\Omega}^0 f^1_{m\gamma} + \tilde{f}^{0\dagger}_{mj\gamma'} \delta\Lambda_{\Omega} \xi^0_{m\gamma} = \lambda^0_m \tilde{f}^{0\dagger}_{mj\gamma'}   f^1_{m\gamma} + \delta\lambda^1_{m\gamma} \tilde{f}^{0\dagger}_{mj\gamma'}  \xi^0_{m\gamma}\,.
\end{equation} 

Since $\tilde{f}^{0 \dagger}_{mj\gamma'}$ is a left eigenvector of $\Lambda^0_{\Omega}$, the first terms in the LHS and RHS cancel out. So we are left with
\begin{equation}
	\tilde{f}^{0\dagger}_{mj\gamma'}  \delta\Lambda_{\Omega} \xi^0_{m\gamma} =   \delta\lambda^1_{m\gamma} \tilde{f}^{0\dagger}_{mj\gamma'} \xi^0_{m\gamma}\,.
\end{equation}
If we decompose back $\xi^0_{m\gamma} = \sum_{\alpha} a_{\gamma \alpha} f^0_{m1\alpha} + \sum_{\beta} b_{\gamma \beta} f^0_{m2\beta}$ we can rewrite this equation as:

\begin{equation}
	\sum_\alpha a_{\gamma \alpha} V^{j1}_{m\gamma'm\alpha} + \sum_\beta b_{\gamma \beta} V^{j2}_{m\gamma'm\beta} = \delta\lambda^1_{m\gamma} \left(a_{\gamma \gamma'}\delta^{j1} + b_{\gamma \gamma'}\delta^{j2}\right)\,.
\end{equation}
where we have used the orthonormality relation $\tilde{f}^{0\dagger}_if^0_j = \delta_{ij}$ and we have introduced the operator
\begin{equation}
	V^{ij}_{m\alpha n\beta} = \tilde{f}^{0 \dagger}_{mi\alpha}  \delta \Lambda_{\Omega} f^0_{nj\beta}\,.
\end{equation}
Because of the block structure of $P_0 \delta \Lambda_{\Omega}$, it is clear that $V^{11}_{m\alpha n\beta} = V^{22}_{m\alpha n\beta} = 0$ while the other components take the following form
\begin{equation}
\begin{aligned}
	V^{12}_{m\alpha n \beta} & = -I(r) \int_A d^3x_a \int_A d^3z_a P(\vec{x}_a,\vec{z}_a) f_{m 1 \alpha}^0(\vec{x}_a)\cdot\\
	& \cdot\int_B d^3z_b \int_B d^3y_b P(\vec{z}_b,\vec{y}_b) f_{n 2 \beta}^0(\vec{y}_b)\\
	& \equiv -I(r) C_{m\alpha n\beta}\,.
\end{aligned}
\end{equation}

Note the symmetry
\begin{equation}
V^{12}_{m\alpha n\beta} = V^{21}_{n\beta m \alpha}\,,	
\end{equation}
which makes the definition of $C_{m\alpha n \beta}$ meaningful. We further define the set of matrices
\begin{equation}
	\left( C_{mn} \right)_{\alpha \beta} \equiv C_{m\alpha n \beta}\,,
\end{equation}
so that the equation for the first order perturbation $\delta \lambda_{m\gamma}^1$ can be rewritten as a block matrix equation
\begin{equation}
	-I(r) \begin{pmatrix}
		0 & C_{mm}\\
		C^T_{mm} & 0\\
	\end{pmatrix} \begin{pmatrix}
		a_{\gamma}\\
		b_{\gamma}
	\end{pmatrix} = \delta \lambda^1_{m\gamma} \begin{pmatrix}
		a_\gamma\\
		b_\gamma
	\end{pmatrix} \,.
\end{equation}

In the case that $\lambda^0_m$ is not a common eigenvalue of $\Lambda^0$ in both regions $A$ and $B$, then either the coefficients $a_\gamma$ or $b_\gamma$ (notice that they are vectors) vanish and so does the perturbation $\delta \lambda^1_m$. On the contrary, if $\lambda^0_m$ is indeed a common eigenvalue, then this equation becomes an eigenvalue equation that is solved by means of a characteristic polynomial
\begin{equation}
\begin{aligned}
	&\det \begin{vmatrix}
		x 1_{M_m \times M_m} & -C_{mm}\\
		-C^T_{mm} & x_{N_m \times N_m}
	\end{vmatrix} \\
	&= \det(x1_{M_m\times M_m})\det \left(x1_{N_m \times N_m}-x^{-1} C^{T}_{mm}C_{mm}\right)\\
	&= x^{M_m-N_n} \det \left(x^2 1_{N_m \times N_m} - C^T_{mm}C_{mm} \right)\,,
\end{aligned}
\end{equation}
where $M_m$ and $N_m$ are the degeneracies of the eigenvalues $\lambda^0_m$ in each region with the convention $M_m \ge N_m$. In other words, the perturbation is linked to the eigenvalue problem for the matrix $C^T_{mm}C_{mm}$ which is a symmetric positive semi-definite matrix, since $C_{mm}$ is real and symmetric. This means that for all its eigenvalues $c_{m\alpha} \ge 0$ and then the perturbation $\delta \lambda^1_m$ either vanishes or comes in pairs of opposite sign
\begin{equation}
	\delta \lambda^1_{m\gamma} = \pm I(r) \sqrt{c_{m\gamma}}\,,
\end{equation}
and thus the first order perturbation to the entropy vanishes because the following combination also vanishes
\begin{equation}
	\sum_\gamma \delta \lambda^1_{m\gamma} \frac{dh}{d\lambda_m} \bigg|_{\lambda_m= \lambda^0_m} = 0\,.
\end{equation}
Next, we need to deal with the second order perturbation. Recall the relevant equation
\begin{equation}
	\Lambda_{\Omega}^0 f^2_{m\gamma} + \delta\Lambda_{\Omega} f^1_{m\gamma} = \lambda^0_m f^2_{m\gamma} + \delta\lambda^1_{m\gamma}f^1_{m\gamma} + \delta\lambda^2_{m\gamma} \xi^0_{m\gamma}\,.
\end{equation}
We can multiply this time my $\tilde{\xi}^{0\dagger}_{mi\gamma'}$ on the left in order to get rid of the first terms of the left- and right-hand side:
\begin{equation}
	\tilde{\xi}_{m\gamma'}^{0\dagger} \delta \Lambda_{\Omega} f^1_{m\gamma} = \tilde{\xi}_{m\gamma'}^{0\dagger} \delta\lambda^1_{m\gamma}f^1_{m\gamma} + \tilde{\xi}_{m\gamma'}^{0\dagger} \delta\lambda^2_{m\gamma} \xi^0_{m\gamma}\,.
\end{equation}
We need an explicit expression for $f^1_{m\gamma}$. Let us look again at the first order perturbation
\begin{equation}
	\Lambda_{\Omega}^0 f^1_{m\gamma} + \delta\Lambda_{\Omega} \xi^0_{m\gamma} = \lambda^0_m f^1_{m\gamma} + \delta\lambda^1_{m\gamma} \xi^0_{m\gamma}\,.
\end{equation}
This means that
\begin{equation}
	f^1_{m\gamma} = \left(\Lambda_{\Omega}^0 - \lambda^0_m \right)^{-1} \left(\delta \lambda_{m\gamma}^1 - \delta\Lambda_{\Omega}\right) \xi^0_{m\gamma}\,.
\end{equation}
We now insert the identity operator
\begin{equation}
\begin{aligned}
	f^1_{m\gamma} &  = \left(\Lambda_{\Omega}^0 - \lambda^0_m \right)^{-1} \left(\sum_{n,j,\alpha} f^0_{nj\alpha} \tilde{f}^{0\dagger}_{nj\alpha}  \right) \left(\delta \lambda_{m\gamma}^1 - \delta\Lambda_{\Omega}\right) \xi^0_{m\gamma}\\
	& = \sum_{n\neq m,j,\alpha} \left(\lambda_m^0 - \lambda^0_n \right)^{-1} f^0_{nj\alpha}\tilde{f}^{0\dagger}_{nj\alpha} \delta\Lambda_{\Omega} \xi^0_{m\gamma}\,.
\end{aligned}
\end{equation}
Note that the addend would vanish if $m=n$ due to the equation for the first-order perturbation. Now we can plug this in the equation for $\delta\lambda^2_{m\gamma}$
\begin{equation}
\begin{aligned}
	&\tilde{\xi}^{0\dagger}_{m\gamma'}\delta \lambda^2_{m\gamma} \xi^0_{m\gamma} \\
& = \tilde{\xi}^{0\dagger}_{m\gamma'} \left(\delta\Lambda_{\Omega} - \delta \lambda^1_{m\gamma} \right) f^1_{m\gamma}\\
	& = \tilde{\xi}^{0\dagger}_{m\gamma'} \left(\delta\Lambda_{\Omega} - \delta \lambda^1_{m\gamma} \right)\sum_{n\neq m,j,\alpha} \left(\lambda_m^0 - \lambda^0_n \right)^{-1} f^0_{nj\alpha} \tilde{f}^{0\dagger}_{nj\alpha} \delta\Lambda_{\Omega} \xi^0_{m\gamma}\\
	& = \sum_{n\neq m,j,\alpha} \left(\lambda_m^0 - \lambda^0_n \right)^{-1} \tilde{\xi}^{0\dagger}_{m\gamma'} \delta\Lambda_{\Omega}   f^0_{nj\alpha} \tilde{f}^{0\dagger}_{nj\alpha} \delta\Lambda_{\Omega} \xi^0_{m\gamma}\,.
\end{aligned}
\end{equation}
In the last line we used $\delta\lambda^1_{m\gamma} \tilde{\xi}^{0 \dagger}_{m\gamma'} f^0_{nj\alpha} = 0$ for $n\neq m$. Finally, since $\tilde{\xi}^{0\dagger}_{m\gamma'}\xi^0_{m\gamma} = \delta_{\gamma \gamma'}$ 
\begin{equation}
\begin{aligned}
	\delta \lambda^2_{m\gamma} & = \!\! \sum_{n\neq m,j,\alpha} \left(\lambda_m^0 - \lambda^0_n \right)^{-1} \tilde{\xi}^{0\dagger}_{m\gamma} \delta\Lambda_{\Omega}   f^0_{nj\alpha} \tilde{f}^{0\dagger}_{nj\alpha} \delta\Lambda_{\Omega} \xi^0_{m\gamma}\\
	& \equiv \sum_{n\neq m} \left(\lambda_m^0 - \lambda^0_n \right)^{-1} \tilde{\xi}^{0\dagger}_{m\gamma} \delta\Lambda_{\Omega}  \hat{\phi}_n \delta\Lambda_{\Omega} \xi^0_{m\gamma}\,,
\end{aligned}
\end{equation}
where we have introduced the projector onto the subspace spanned by the eigenvectors with eigenvalue $\lambda_n$
\begin{equation}
		\hat{\phi}_n = \sum_{j,\alpha} f^0_{nj\alpha} \tilde{f}^{0\dagger}_{nj\alpha}\,.
\end{equation}
Now we compute the perturbation to the entropy due to the second order perturbation $\delta \lambda^2_{m\gamma}$
\begin{equation}
\begin{aligned}
	&\sum_{m,\gamma} \delta \lambda^2_{m\gamma} \frac{dh}{d\lambda_m} \bigg|_{\lambda_m=\lambda_m^0}\\
	& = \sum_{m,\gamma}\sum_{n\neq m} \left(\lambda_m^0 - \lambda^0_n \right)^{-1} \tilde{\xi}^{0\dagger}_{m\gamma} \delta\Lambda_{\Omega}  \hat{\phi}_n \delta\Lambda_{\Omega} \xi^0_{m\gamma} \frac{dh}{d\lambda_m} \bigg|_{\lambda_m=\lambda_m^0}\\
	& = \sum_m \sum_{n\neq m} \left(\lambda_m^0 - \lambda^0_n \right)^{-1}  \textrm{Tr} \left( \hat{\phi}_m \delta\Lambda_{\Omega} \hat{\phi}_n \delta\Lambda_{\Omega} \right)\frac{dh}{d\lambda_m} \bigg|_{\lambda_m=\lambda_m^0}\\
	& = \sum_n \sum_{m > n} \left(\lambda_m^0 - \lambda^0_n \right)^{-1}  \textrm{Tr} \left( \hat{\phi}_m \delta\Lambda_{\Omega} \hat{\phi}_n \delta\Lambda_{\Omega} \right) \times \\
	&\quad \times \left(\frac{dh}{d\lambda_m} \bigg|_{\lambda_m=\lambda_m^0} - \frac{dh}{d\lambda_n} \bigg|_{\lambda_n=\lambda_n^0} \right)\,.
\end{aligned}
\end{equation}
In the last line we simply relabelled the indices so that $m > n$. Furthermore, the alternative expression for the projector was used
\begin{equation}
	\sum_\gamma \xi^0_{m\gamma} \tilde{\xi}^{0\dagger}_{m\gamma} = \hat{\phi}_m \,.
\end{equation}
What is the sign of this expression? Let us take a look at the trace
\begin{equation}
\begin{aligned}
	& \textrm{Tr} \left( \hat{\phi}_m \delta\Lambda_{\Omega} \hat{\phi}_n \delta\Lambda_{\Omega} \right) \\
	& = \sum_{i,\alpha,j,\beta} \left(\tilde{f}^{0\dagger}_{ni\alpha} \delta\Lambda_{\Omega} f^0_{mj\beta} \right)\left( \tilde{f}^{0\dagger}_{mj\beta} \delta\Lambda_{\Omega} f^0_{ni\alpha} \right) =\\
	& = \sum_{i,\alpha,j,\beta} V^{ij}_{n\alpha m\beta} V^{ji}_{m\beta n\alpha} = \sum_{i,\alpha,j,\beta} \left(V^{ij}_{n\alpha m\beta} \right)^2\\
	& = \sum_{\alpha \beta} I(r)^2 \left(C_{n\alpha m \beta}\right)^2 \ge 0 \,.
\end{aligned}
\end{equation}

It's time to compute the derivatives of $h$
\begin{itemize}
	\item Function
	\begin{equation}
	\begin{aligned}
		h(\lambda) = &  \left(\sqrt{\lambda}+1/2\right)\log\left( \sqrt{\lambda}+1/2\right)\\
		& - \left(\sqrt{\lambda}-1/2\right)\log\left( \sqrt{\lambda}-1/2\right) \,.
	\end{aligned}
	\end{equation}
	
	\item First derivative
	\begin{equation}
	\begin{aligned}
		\frac{dh}{d\lambda} (\lambda) = \frac{1}{2\sqrt{\lambda}} \left[ \log\left(\sqrt{\lambda}+1/2\right) - \log\left(\sqrt{\lambda}-1/2\right) \right]\\
		> 0
		\quad \textrm{for} \quad \lambda > 1/4 \,.
	\end{aligned}
	\end{equation}
	\item Second derivative
	\begin{equation}
	\begin{aligned}
		\frac{d^2 h}{d\lambda^2} (\lambda) = \frac{\frac{4 \sqrt{\lambda }}{1-4 \lambda }+\log \left(\sqrt{\lambda
   }-\frac{1}{2}\right)-\log \left(\sqrt{\lambda }+\frac{1}{2}\right)}{4 \lambda ^{3/2}}\\ < 0
   \quad \textrm{for} \quad \lambda> 1/2 \,.
    \end{aligned}
	\end{equation}
\end{itemize}

Furthermore, the first derivative is positive but monotonically decreasing, while the second derivative is negative but monotonically increasing. Both tend to 0 for large $\lambda$ and blow up for  $\lambda \rightarrow 1/4$

In particular, if $m > n$ then $\lambda_m > \lambda_n$ and so
\begin{equation}
	\left(\frac{dh}{d\lambda_m} \bigg|_{\lambda_m=\lambda_m^0} - \frac{dh}{d\lambda_n} \bigg|_{\lambda_n=\lambda_n^0} \right) < 0\,,
\end{equation}
and the sign of the perturbation is non-positive.

There is also a second order perturbation coming from the term
\begin{equation}
	\sum_{m\gamma}\left(\delta\lambda^1_m\right)^2 \frac{d^2h}{d\lambda^2_m} \bigg|_{\lambda_m=\lambda_m^0} \le 0\,,
\end{equation}
and thus the sign of this perturbation is non-positive as well.

The last step is to plug everything into the formula for the mutual information between the two regions
\begin{equation}
\begin{aligned}
	I(A,B) & = S_A + S_B - S_{AB}  \\
	& =  -\sum_i \left[\delta\lambda_i \frac{dh}{d\lambda_i}\bigg|_{\lambda_i=\lambda_i^0} + \frac{1}{2}(\delta\lambda_i)^2 \frac{d^2h}{d\lambda_i^2}\bigg|_{\lambda_i=\lambda_i^0}\right]\\
	& = - I(r)^2 G \left(A,B\right) \ge 0\,.
\end{aligned}
\end{equation}
Then there is a non-negative mutual information between disjoin regions that is enhanced due to inflation. Here $G(A,B)$ is a function of the size and possibly the shape of the regions $A$ and $B$, e.g. for two spherical regions of radii $R_1$ and $R_2$ we would have $G(A,B) = G(R_1,R_2)$, but its precise form is not that easy to compute.

Nevertheless, $G(R_1,R_2)$ is a function of the short-range behavior of the operator $P$ and as such its leading term is expected to agree with the Minkowski computation. In that case, one has the following result for the mutual information\cite{Shiba:2010dy}
\begin{equation}
    I_M(A,B) = - \frac{1}{16 \pi^4 \,r^4} G(A,B) \,.
    \label{eq:IM}
\end{equation}

Notice that we use the convention of factoring out of $G(A,B)$ not only the long-range dependence on $r$ but also numerical coefficients coming from $X(\vec{x},\vec{y}$). The function $G(A,B)$ was computed numerically by Shiba in \cite{Shiba:2012np} and found
\begin{equation}
	G(R_1,R_2) \simeq - \frac{1}{4} R_1^2 R_2^2 \times 16 \pi^4\,.
\end{equation}
We take this computation to be valid in leading order for our case, because the kernel $X(x,y)$ is equal to the Minkowski kernel for most momenta. Dimensions agree but notice that $R_i$ are comoving, not physical, radii. Then we arrive to the result

\begin{equation}
\begin{aligned}
    I(A,B) \simeq & \frac{1}{4} I(r)^2 R_1^2 R_2^2 \times 16\pi^4\\ 
     \simeq & \frac{1}{16} \left( \frac{\eta}{\eta_{\text{end}}} \right)^4 \frac{R_1^2R_2^2}{\eta_{\rm end}^4} \left[1-\gamma+\log\left(\frac{-\eta_0}{r}\right)\right. \\& \left. + {\rm Ci}\left(\frac{r}{\eta}\right) 
    -\left(\frac{\eta}{r}\right)
    \sin\left(\frac{r}{\eta}\right)
    \right]^2 \\ 
\end{aligned}
\label{eq:Ien1}
\end{equation}

Where we have used the approximation from eq. (\ref{eq:approx}), which is valid for $r < -\eta_0$. More compactly, we arrive at

\begin{equation}
    I(A,B) \simeq \frac{1}{16} \frac{R_1^2R_2^2} {\eta_{\rm end}^4} \left( \frac{\eta}{\eta_{\text{end}}} \right)^4  \left[1-\gamma+\log\left(\frac{-\eta_0}{r}\right)\right]^2\,,
\label{eq:Ien2}
\end{equation}

using the approximation from eq. (\ref{eq:approx2}), which is valid for $ - \eta_0 > r > \eta$. The long-range behavior is inherited by the mutual information and thus an enhancement is obtained due to inflation. On the one hand, it decays logarithmically and, therefore, slower than inverse powers of $r$. On the other hand, the ratio $R^2_1 R_2^2 / \eta_{\rm end}^4$ can be potentially very large and does not depend on the distance, as opposed to the mutual information for the Minkowski vacuum, which behaves as $R_1^2 R_2^2 / r^4$, which is necessarily small for the perturbative approach to work. Furthermore, it continues growing with time as $\eta^4/\eta_{\rm end}^4$ during the radiation era.

This result is valid for super-horizon scales during the radiation era. It can also applied before the radiation era, during inflation, by setting $\eta = \eta_{\rm end}$ and replacing the remaining $\eta_{\rm end}^{-4}$ factor by the same power of the conformal time at which it is computed.

\begin{figure}
    \centering
    \includegraphics[width=\linewidth]{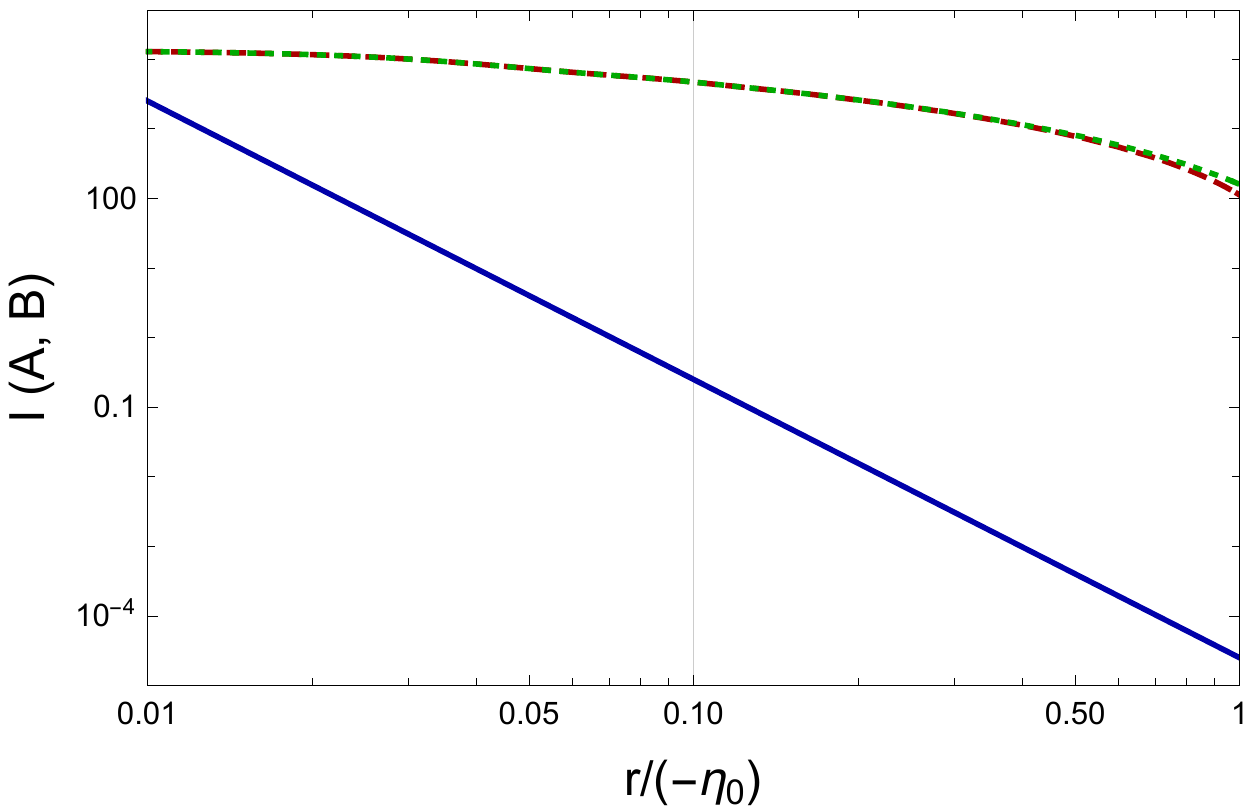}
    \caption{Comparison between the mutual information of the Minkowski vacuum Eq.~(\ref{eq:IM}) (blue line) and the enhanced mutual information Eq.~(\ref{eq:Ien1},\ref{eq:Ien2}) (red dashed and green dotted lines, as in Fig.~2), for $\eta_{\rm end} = -0.1$, $\eta_0 = -10$, $\eta = \eta_{\rm end}$ and $R_1 = R_2 = 1$ (gray vertical line).    }
    \label{fig:info}
\end{figure}

\section{VI. Phenomenology}

Mutual information behaves differently from other thermodynamical quantities during the radiation era. For instance, the mutual information between two regions of fixed physical size $R$ remains constant, since $a \sim \eta$,  as opposed to other quantities that dilute due to the expansion of the universe, such as the energy density of relativistic species.

The enhanced mutual information seems intuitively to be  connected to some of the main predictions of inflation, such as the leading homogeneous and isotropic nature of the universe and the common causal past of the observable universe. For instance, the CMB temperature anisotropies are characterized with the 2-point correlation function of curvature perturbations. The enhanced mutual information offers a new perspective on a well-known fact, namely that fluctuations in distant points in the sky are tightly related. Quantum correlations in the CMB have been explored by computing the quantum discord of primordial perturbations in momentum space \cite{Martin:2015qta}. The enhanced mutual information is a first step towards a similar study of quantum correlations in position space in the CMB and possibly other cosmological observables. 

Following the ideas presented in \cite{Espinosa-Portales:2019peb} we state that, should certain regions collapse to form Primordial Black Holes (PBH) during the radiation era, the PBH will inherit the enhanced mutual information by the collapsing regions. Whether these signals enhanced genuinely quantum correlations between the PBH remains an open question, since mutual information is a measure of both classical and quantum correlations. 

Of course, in our computation we considered a toy model for inflation that delivers an exactly flat power spectrum. Under such circumstances the formation of a PBH is an extremely unlikely event. Hence, in order to compute the mutual information between two PBH, we would need to consider the power spectrum of the particular inflationary model leading to sufficiently abundant PBH formation~\cite{GarciaBellido:1996qt}. It deviates from flatness at scales comparable to the comoving size of the PBH at formation time (or, equivalently, the size of the Hubble scale at formation time), but not for scales well-probed such as the CMB scales. This should make no difference for the mutual information shared by PBH separated by distances so large that the power spectrum at the corresponding scale is flat or nearly flat. For those we can state their pair-wise mutual information at formation time to be given by

\begin{equation}
	I_{\rm PBH} \simeq \frac{1}{16} \frac{R_1^2R_2^2}{\eta_{\rm end}^4} \left( \frac{\eta}{\eta_{\text{end}}} \right)^4 \left[\log\left(\frac{-\eta_0}{r}\right)+1-\gamma \right]^2\,.
\end{equation}

This mutual information characterizes the properties of the  network that PBH form, at least at formation time and for PBH at large enough distances. One may find even larger values for PBH in dense clusters, once they find eachother and merge. We leave for future work the application of this methodology to particular inflationary models and PBH formation scenarios.

Entangled Black Holes have been considered previously in the literature, for instance in the context of the celebrated ER = EPR correspondence \cite{Maldacena:2013xja}. In this framework, one could picture the network of entangled PBH as a network of black holes connected by wormholes that fill the entire Universe. In that case, the mutual information shared by the PBH would most likely be relevant in order to characterize the wormholes that connect them, as long as genuinely quantum correlations are enhanced as well. For instance, two black holes connected by an Einstein-Rosen bridge would be maximally entangled in the ER = EPR correspondence, and so their mutual information is maximal and equal to the Bekenstein entropy of a single Black Hole.

We wonder whether the entropy of the PBH network can be interpreted as thermodynamical entropy and, in that case, lead to some kind of entropic forces that would affect the dynamics of the network. We leave this discussion for future work.

\section{VII. Conclusions}

The quantum origin of primordial curvature perturbations generated during inflation has provided a fascinating explanation for the origin of the matter distribution on large scales. However, it is often thought to offer no distinctive signature or observational feature compared to simply postulating the existence of a classical Gaussian (free) stochastic field of density perturbations. This is due to the suppression of the decaying mode thanks to squeezing, a phenomenon called \textit{decoherence without decoherence}~\cite{Polarski:1995jg}, which is actually necessary in order to reproduce the apparently classical features of the primordial power spectrum of matter fluctuations seen in the CMB and LSS.

Nevertheless, there has been  recent interest on the quantum nature of the matter distribution and how to properly distinguish quantum from classical perturbations. Although the decaying mode is hopelessly suppressed in both slow-roll and ultra-slow-roll inflation~\cite{dePutter:2019xxv}, there are actually features of the primordial bi-spectrum (the 3-point correlation function) that would be distinctively quantum and may be probed in the future~\cite{Green:2020whw}. On the other hand, the quantum nature of inflationary fluctuations can be explored with rare but highly non-linear phenomena like primordial black hole collapse during the radiation era, that arises precisely because of large non-Gaussian tails due to quantum diffusion during inflation~\cite{Ezquiaga:2019ftu}. These events could provide the best clue as to the quantum nature of matter fluctuations generated during inflation, affecting structure formation and constituting a significant component of dark matter~\cite{Garcia-Bellido:2019tvz}. We believe the importance of the quantum origin of cosmological perturbations should not be understated.

In this paper we have studied the mutual information between two disjoint regions at super-horizon scales in a radiation-dominated universe filled with curvature perturbations of inflationary origin. This enhanced mutual information has a quantum origin, in the sense that is linked to squeezing and particle creation, and may be linked to genuine quantum correlations. Future research will be required to establish this. 

Even if the enhanced mutual information is dominated by classical correlations, our results offer a new approach to the predictions of inflation, as it is related to a scale-invariant power spectrum of primordial perturbations. Furthermore, future research in the topic of entropic forces, which has precedence in cosmology, could provide relevant observational features.

\section{Acknowledgements}

We thank Germ\'an Sierra for his insightful comments. The authors acknowledge support from the Spanish Research Project PGC2018-094773-B-C32 (MINECO-FEDER) and the Centro de Excelencia Severo Ochoa Program SEV-2016-0597. The work of LEP is funded by a fellowship from "La Caixa" Foundation (ID 100010434) with fellowship code LCF/BQ/IN18/11660041 and the European Union Horizon 2020 research and innovation programme under the Marie Sklodowska-Curie grant agreement No. 713673.

\section{Appendix. Beyond the random phase approximation}

One may wonder whether assuming that the squeezing phases $\delta_k$ are random has a noticeable effect on the mutual information of primordial perturbations. In the following we argue why it is not the case.

In the case that $\braket{vp + pv} \neq 0$ then one cannot simply compute the entropy of the quantum state by finding the eigenvalues of the operator $\Lambda$, as described in section IV. Instead, one needs to consider an operator built from the larger field:

\begin{equation}
    \chi = \begin{pmatrix}
        v \\
        \pi
    \end{pmatrix}
\end{equation}

Its 2-point correlation function contains all the 2-point correlation functions of the state:

\begin{equation}
    \Delta(\vec{x},\vec{y}) = \frac{1}{2} \braket{\chi(\vec{x}) \chi(\vec{y}} = \begin{pmatrix}
        X(\vec{x},\vec{y}) & \frac{1}{2}C(\vec{x},\vec{y})\\
        \frac{1}{2}C(\vec{x},\vec{y} & P(\vec{x},\vec{y})
    \end{pmatrix}
\end{equation}

Where:

\begin{equation}
    C(\vec{x},\vec{y}) = \braket{v(\vec{x}) p(\vec{y}) + p(\vec{y})v(\vec{x})}
\end{equation}

This correlation function transforms under symplectic transformations as:

\begin{equation}
    \Delta \rightarrow S \Delta S^{\dagger}
\end{equation}

Such transformations are not similarity transformation and, hence, do not leave the eigenvalues of $\Delta$ invariant. Still, Williamson's theorem guarantees that there exists a symplectic transformation that brings $\Delta$ to a diagonal form \cite{Berges:2017hne}. Note that any symplectic transformation $S$ preserves the symplectic form:

\begin{equation}
	S \Omega S^{\dagger} = \Omega \quad \textrm{where} \quad
	\Omega = \begin{pmatrix}
		0 & i \\
		-i & 0\\
	\end{pmatrix}
\end{equation}

Or, equivalently:

\begin{equation}
	S^{\dagger} = \Omega S^{-1} \Omega
\end{equation}

This means that the problem of finding symplectic eigenvalues of $\Delta$ is equivalent to finding conventional eigenvalues of $\Delta \Omega$

\begin{equation}
	\Delta \Omega = \begin{pmatrix}
		-\frac{i}{2} C & i X \\
		-i P & \frac{i}{2} C\\
	\end{pmatrix}
\end{equation}

If we assume random phases, then $C=0$ and the eigenvalues of $\Delta \Omega$ are those of $\sqrt{\Lambda}$, so that both the formalism used in section IV and the one presented here are consistent. If $C\neq 0$, we need to study the eigenvalue problem of this operator. The determinant of a block matrix admits the following decomposition
\begin{equation}
\begin{aligned}
	&M = \begin{pmatrix}
		M_{11} & M_{12} \\
		M_{21} & M_{22} \\
	\end{pmatrix}\\[3mm]
	&\textrm{det}(M) = \textrm{det}(M_{22}) \textrm{det} (M_{11} - M_{12} M_{22}^{-1} M_{21})\,.
\end{aligned}
\end{equation}

We are interested in the determinant of $\Delta \Omega - \lambda$ in order to find the eigenvalues of $\Delta \Omega$ and thus
\begin{equation}
\begin{aligned}
	&\textrm{det} (\Delta \Omega - \lambda) =\\
	&\textrm{det} \left(\frac{i}{2}C - \lambda \right) \textrm{det} \left( \frac{i}{2}C - \lambda - X \left(\frac{i}{2}C - \lambda \right)^{-1} P  \right)\,.
\end{aligned}
\end{equation}
This expression admits two approximations. First, since we are interested in the perturbative regime, $C$ will have a subdominant contribution in the first determinant and can be neglected. This argument is valid as well for the first term of the second determinant. Second, since $C$ deals with only a subset of momentum modes, we can assume that it has a norm smaller than that of the identity and hence we can expand the inverse as
	\begin{equation}
		- \lambda^{-1} \left(1 - \lambda^{-1} \frac{i}{2}C \right)^{-1} \simeq - \lambda^{-1} \left( 1 + \lambda^{-1} \frac{i}{2}C \right)\,,
	\end{equation}
and thus
\begin{equation}
	\textrm{det}(\Delta \Omega - \lambda) \simeq \textrm{det}\left[ - \lambda + \lambda^{-1} X (1 + i\lambda^{-1} C/2  ) P \right] \,.
\end{equation}
Recall that the dominant perturbative contribution is given by $X$ being perturbative and $P$ being non-perturbative. Then, $C$ will be non-perturbative as well. Since it is only non-vanishing for momenta affeced by inflation, it does not affect the more relevant high-momentum modes of $P$. Hence, we can conclude than the effect of averaging over the squeezing phases has a negligible effect on the mutual information of primordial perturbations.

\bibliographystyle{h-physrev}
\bibliography{paperMI}

\end{document}